\documentclass{ws-ijgmmp}
\usepackage{amsmath}
\usepackage{amssymb}
\numberwithin{equation}{section}

\begin{document}
\markboth{G. Esposito, E. Battista, E. Di Grezia}
{Bicharacteristics and Fourier integral operators in Kasner spacetime}

\title{BICHARACTERISTICS AND FOURIER INTEGRAL OPERATORS IN KASNER SPACETIME}

\author{GIAMPIERO ESPOSITO}
\address{Istituto Nazionale di Fisica Nucleare, Sezione di Napoli,
Complesso Universitario di Monte S. Angelo, \\
Via Cintia Edificio 6, 80126 Napoli, Italy \\
\email{gesposit@na.infn.it}}

\author{EMMANUELE BATTISTA}
\address{Dipartimento di Fisica, Complesso Universitario di Monte S. Angelo, \\
Via Cintia Edificio 6, 80126 Napoli, Italy \\
Istituto Nazionale di Fisica Nucleare, Sezione di Napoli,
Complesso Universitario di Monte S. Angelo, \\
Via Cintia Edificio 6, 80126 Napoli, Italy \\
\email{ebattista@na.infn.it}}

\author{ELISABETTA DI GREZIA}
\address{Istituto Nazionale di Fisica Nucleare, Sezione di
Napoli, Complesso Universitario di Monte S. Angelo, \\ 
Via Cintia Edificio 6, 80126 Napoli, Italy \\
\email{digrezia@na.infn.it}}

\maketitle

\begin{abstract}
The scalar wave equation in Kasner spacetime is solved, first for a particular choice of Kasner parameters, by  
relating the integrand in the wave packet to the Bessel functions. 
An alternative integral representation is also displayed, which relies upon the method of integration
in the complex domain for the solution of hyperbolic equations with variable coefficients.
In order to study the propagation of wave fronts, we integrate the equations of bicharacteristics
which are null geodesics, and we are able to express them, for the first time in the literature,
with the help of elliptic integrals for another choice of Kasner parameters.
For generic values of the three Kasner parameters, the solution of the Cauchy problem is built through a pair
of integral operators, where the amplitude and phase functions in the integrand solve a coupled system
of partial differential equations. The first is the so-called transport equation, whereas the second is
a nonlinear equation that reduces to the eikonal equation if the amplitude is a slowly varying function. 
Remarkably, the analysis of such a coupled system is proved to be equivalent to building first an auxiliary
covariant vector having vanishing divergence, while all nonlinearities are mapped into solving a 
covariant generalization of the Ermakov-Pinney equation for the amplitude function. Last, from a linear
set of equations for the gradient of the phase one recovers the phase itself.
This is the parametrix construction that relies upon Fourier-Maslov integral operators,
but with a novel perspective on the nonlinearities in the dispersion relation.
Furthermore, the Adomian method for nonlinear partial differential equations is applied to
generate a recursive scheme for the evaluation of the amplitude function in the parametrix.
The resulting formulas can be used to build self-dual solutions to the field equations of 
noncommutative gravity, as has been shown in the recent literature. 
\end{abstract}


\section{Introduction}

Although it is well known that the wavelike phenomena of classical physics are ruled by hyperbolic
equations, there are at least two modern motivations for studying the scalar wave equation on curved
Lorentzian four-manifolds. They are as follows.
\vskip 0.3cm
\noindent
(i) In the course of studying the Einstein vacuum equations in four dimensions, i.e.
\begin{equation}
R_{\alpha \beta}-{1 \over 2}g_{\alpha \beta}R=0 \Longrightarrow R_{\alpha \beta}=0,
\label{(1.1)}
\end{equation}
it was conjectured in Ref. \cite{Kleinerman} that they admit local Cauchy developments for 
initial data\footnote{The Cauchy problem consists in finding a Lorentzian four-metric $g$ 
satisfying Eq. (1.1) such that the metric $h$ induced by $g$ on a given spacelike hypersurface 
$\Sigma_{0}$ and the extrinsic-curvature tensor $k$ of $\Sigma_{0}$ are prescribed.} sets
$(\Sigma_{0},h,k)$ with locally finite $L^{2}$ curvature and locally finite $L^{2}$ norm of
the first covariant derivatives of $k$. This means that the spacetime constructed by evolution
from smooth data can be smoothly continued, together with a time foliation, as long as the curvature
of the foliation and the first covariant derivatives of its extrinsic curvature remain
$L^{2}$-bounded on the leaves of the foliation \cite{K2013}. The proof that this is indeed the case
relies on a number of technical ingredients, including the construction of a parametrix (an
approximate Green function of the wave operator, that provides a progressive wave representation
\cite{Treves} for solutions of the wave equation) 
for solutions of the homogeneous wave equation
\begin{equation}
\Box_{g}\phi=g^{\mu \nu}\nabla_{\mu}\nabla_{\nu}\phi=0
\label{(1.2)}
\end{equation}
on a fixed Einstein vacuum background. One has then to obtain control of the parametrix and of its
error term by using only the fact that the curvature tensor is bounded in $L^{2}$ \cite{Szeftel}.
Note that, at a deeper level, the metric $g$ can be viewed to determine the elliptic
or hyperbolic nature\footnote{If $M$ is a connected, four-dimensional, Hausdorff four-manifold 
of class $C^{\infty}$, a linear partial differential  operator is a linear map
$$
L: u \in C^{\infty}(M) \rightarrow (Lu) \in C^{k}(M),
$$
with coefficients $a^{i_{1}...i_{m}}$ given by functions of class $C^{k}$.
The {\it characteristic polynomial} of the operator $L$ at a point $x \in M$ is
$$
H(x,\xi)=\sum a^{i_{1}...i_{m}}(x)\xi_{i_{1}}... \xi_{i_{m}},
$$
where $\xi_{i}$ is a cotangent vector at $x$. The cone in the cotangent plane $T_{x}^{*}$ at $x$ 
defined by
$$
H(x,\xi)=0
$$
is called the characteristic cone (or conoid). By construction, such a cone is independent
of the choice of coordinates, because the higher order terms
(also called leading or principal symbol) of $L$ transform into
higher-order terms by a change of coordinates. The operator $L$ is said to be hyperbolic at $x$
if there exists a domain $\Gamma_{x}$, a convex open cone in $T_{x}^{*}$, such that every
line through $\lambda \in \Gamma_{x}$ cuts the characteristic cone in $m$ real distinct points.
In particular, second-order differential operators with higher-order terms
$$
g^{\alpha \beta}(x){\partial \over \partial x^{\alpha}}
{\partial \over \partial x^{\beta}}
$$
are hyperbolic at $x$ if and only if the cone defined by
$$
H_{2}(x,\xi) \equiv g^{\alpha \beta}(x)\xi_{\alpha}\xi_{\beta}=0
$$
is convex, i.e., if the quadratic form $H_{2}(x,\xi)$ has signature $(1,n-1)$.} 
of the operator $g^{\mu \nu}\nabla_{\mu} \nabla_{\nu}$, where $\nabla_{\mu}$ 
can denote covariant differentiation with respect to the Levi-Civita connection on spacetime,
or on a vector bundle over spacetime, depending on our needs. When $g$ is Riemannian, i.e.
positive-definite, this operator is minus the Laplacian, whereas if $g$ is Lorentzian, one gets the wave
operator. Note also that, in four-dimensional manifolds, our Lorentzian world lies in between two other 
options, i.e. a Riemannian metric $g$ with signature $4$ and elliptic operator 
$g^{\mu \nu}\nabla_{\mu}\nabla_{\nu}$, and a
ultrahyperbolic metric $g$ with signature $0$ and ultrahyperbolic 
operator $g^{\mu \nu}\nabla_{\mu}\nabla_{\nu}$. 
In the so-called Euclidean (or Riemannian) framework used 
by quantum field theorists in functional integration, where the metric
is positive-definite, the most fundamental differential operator is however the Dirac operator, obtained by
composition of Clifford multiplication with covariant differentiation. Its leading symbol is therefore
Clifford multiplication, and it generates all elliptic symbols on compact Riemannian manifolds
\cite{Esposito98}. This reflects the better known property according to which, out of the Dirac operator
and its (formal) adjoint, one can define two operators of Laplace type, as well as powers of these
operators.  
\vskip 0.3cm
\noindent
(ii) Recent work on the self-dual road to noncommutative gravity with twist has found it useful to start from a classical,
undeformed spacetime which is a self-dual solution of the vacuum Einstein equation, e.g. a Kasner spacetime
\cite{DEV14}. Within that framework, it is of interest to solve first the scalar wave equation in such a Kasner
background. Since such a task was only outlined in Ref. \cite{DEV14}, we find it appropriate to develop a systematic
calculus in the present paper.

Relying in part upon Ref. \cite{DEV14}, we begin by considering the scalar wave equation (1.2) 
for a classical scalar field $\phi$ when the Kasner\footnote{With this particular choice of parameters,
we are working on an edge of Minkowski spacetime, i.e., Rindler space. The literature on quantum field
theory and accelerated observers has considered in detail such a space, but in our paper the emphasis
is on partial differential equations in classical physics, hence we do not strictly need ideas
from quantum physics. It would have been helpful to derive the world function in Kasner spacetime
(Appendices B and C) from the world function in Minkowski spacetime, and similarly for the Green function
of Hadamard type, but we have been unable to achieve this. Hence we have limited ourselves to
strict use of Kasner geometry.}
parameters $p_{1} , p_{2} , p_{3}$ take the values $1,0,0$, respectively, i.e.
\begin{equation}
\left(-{\partial^2 \over \partial t^2} -{1 \over t} {\partial \over \partial t}
+{1 \over t^{2}} {\partial^2 \over \partial x^2}
+ {\partial^2 \over \partial y^2} + {\partial^2 \over \partial z^2} \right)\phi=0,
\label{(1.3)}
\end{equation}
where $\phi$ admits the integral representation
\begin{equation}
\phi(t,x,y,z)=\int_{-\infty}^{\infty}{\rm d}\xi_{1} \int_{-\infty}^{\infty}{\rm d}\xi_{2}
\int_{-\infty}^{\infty}{\rm d}\xi_{3} \; A(\xi_{1},\xi_{2},\xi_{3},t)
{\rm e}^{{\rm i}(\xi_{1}x+\xi_{2}y+\xi_{3}z)}.
\label{(1.4)}
\end{equation}
One can then set \cite{DEV14}
\begin{equation}
A(\xi_{1},\xi_{2},\xi_{3},t)={1 \over \sqrt{t}} W(\xi_{1},\xi_{2},\xi_{3},t),
\label{(1.5)}
\end{equation}
where $W(\xi_{1},\xi_{2},\xi_{3},t)$ has to solve, for consistency, the equation \cite{DEV14}
\begin{equation}
\left[{\partial^2 \over \partial t^2}+{\left({1 \over 4}+(\xi_{1})^{2}\right)\over t^{2}}
+(\xi_{2})^{2}+(\xi_{3})^{2} \right]W(\xi_{1},\xi_{2},\xi_{3},t)=0,
\label{(1.6)}
\end{equation}
and the ${1 \over \sqrt{t}}$ term in the factorization (1.5) ensures that, in Eq. (1.6), the first derivative of $W$ is
weighed by a vanishing coefficient. This is a sort of canonical form of linear second-order ordinary differential equations 
with variable coefficients (see Section 10.2 of Ref. \cite{WW27}), 
and Eq. (1.6) can be viewed as a $3$-parameter family of such equations, the parameters being
the triplet $\xi_{1},\xi_{2},\xi_{3}$.

Section II relates Eq. (1.6) to the Bessel functions, studies a specific choice 
of Cauchy data and eventually solves Eq. (1.3) through an integral representation that relies upon
integration in the complex domain. Section III evaluates the bicharacteristics in Kasner spacetime,
relating them to elliptic integrals, while Sec. IV builds the parametrix of our scalar wave equation 
through a pair of integral operators where the integrand consists of amplitude and phase functions.
Concluding remarks and open problems are presented in Sec. V, while relevant background material is
described in the Appendices.

\section{Solving the wave equation with Kasner parameters $(1,0,0)$}

\subsection{Relation with Bessel functions}

From now on, we therefore study until the end of next subsection the ordinary differential equation
\begin{equation}
\left[{{\rm d}^{2}\over {\rm d}t^{2}}+{\left({1\over 4}+(\xi_{1})^{2}\right)\over t^{2}}
+(\xi_{2})^{2}+(\xi_{3})^{2}\right]W(t)=0.
\label{(2.1)}
\end{equation}
This is a particular case of the differential equation
\begin{equation}
\left[{{\rm d}^{2}\over {\rm d}t^{2}}+{(1-2 \alpha)\over t}{{\rm d}\over {\rm d}t}
+\beta^{2}+{(\alpha^{2}-\nu^{2})\over t^{2}}\right]f(t)=0,
\label{(2.2)}
\end{equation}
which is solved by the linear combination
\begin{equation}
f(t)=C_{1}t^{\alpha}J_{\nu}(\beta t)+C_{2}t^{\alpha}Y_{\nu}(\beta t).
\label{(2.3)}
\end{equation}
By comparison of Eqs. (2.1) and (2.2) we find
\begin{equation}
\alpha={1 \over 2}, \; \beta= \sqrt{(\xi_{2})^{2}+(\xi_{3})^{2}}, \;
\nu={\rm i} \xi_{1},
\label{(2.4)}
\end{equation}
and hence, in light of what we pointed out at the end of Sec. I, our partial differential equation (1.4) is solved by
replacing $C_{1}$ and $C_{2}$ in (2.3) by some functions 
$Z_{1}(\xi_{1},\xi_{2},\xi_{3})$ and $Z_{2}(\xi_{1},\xi_{2},\xi_{3})$,
whose form depends on the choice of Cauchy data, i.e. (see Sec. III)
\begin{equation}
W(\xi_{1},\xi_{2},\xi_{3},t)=Z_{1}(\xi_{1},\xi_{2},\xi_{3})
\sqrt{t}J_{{\rm i}\xi_{1}}(t \sqrt{(\xi_{2})^{2}+(\xi_{3})^{2}})
+Z_{2}(\xi_{1},\xi_{2},\xi_{3})\sqrt{t}Y_{{\rm i}\xi_{1}}(t \sqrt{(\xi_{2})^{2}+(\xi_{3})^{2}}).
\label{(2.5)}
\end{equation}
The Bessel function $Y_{{\rm i}\xi_{1}}$ is not regular at $t=0$ and hence, by using this representation, we
are considering an initial time $t_{0} >0$. We use the linearly independent Bessel functions 
$J_{{\rm i}\xi_{1}}$ and $Y_{{\rm i}\xi_{1}}$ which describe accurately the time
dependence of the integrand in Eq. (1.4). Note that the three choices
$$
(p_{1}=1,p_{2}=p_{3}=0), \;
(p_{2}=1,p_{1}=p_{3}=0), \;
(p_{3}=1,p_{1}=p_{2}=0)
$$
are equivalent, since the three coordinates $x,y,z$ in the scalar wave equation \cite{DEV14} are on 
equal footing. Only the calculational details change. More precisely, on choosing
$p_{3}=1,p_{1}=p_{2}=0$,  one finds
$$
\beta=\sqrt{(\xi_{1})^{2}+(\xi_{2})^{2}}, \; \nu={\rm i}\xi_{3},
$$
whereas, upon choosing $p_{2}=1,p_{1}=p_{3}=0$, one finds
$$
\beta=\sqrt{(\xi_{1})^{2}+(\xi_{3})^{2}}, \; \nu={\rm i}\xi_{2}.
$$

\subsection{Role of Cauchy data}

The task of solving our wave equation (1.3) can be accomplished provided that one knows the Cauchy data
\begin{equation}
\Phi_{0} \equiv \phi(t_{0},x,y,z), \; \Phi_{1} \equiv {\partial \phi \over \partial t}(t=t_{0},x,y,z).
\label{(2.6)}
\end{equation}
Indeed, from our Eqs. (1.4), (1.5) and (2.5), 
one finds (denoting by an overdot the partial derivative with respect to $t$)
\begin{eqnarray}
\; & \; & A(\xi_{1},\xi_{2},\xi_{3},t_{0})=(2 \pi)^{-3} 
\int_{-\infty}^{\infty} {\rm d}x 
\int_{-\infty}^{\infty} {\rm d}y 
\int_{-\infty}^{\infty} {\rm d}z \; \Phi_{0}
{\rm e}^{-{\rm i}(\xi_{1}x+\xi_{2}y+\xi_{3}z)} \nonumber \\
&=& Z_{1}(\xi_{1},\xi_{2},\xi_{3})
J_{{\rm i}\xi_{1}}(t_{0}\sqrt{(\xi_{2})^{2}+(\xi_{3})^{2}})
+Z_{2}(\xi_{1},\xi_{2},\xi_{3})
Y_{{\rm i}\xi_{1}}(t_{0}\sqrt{(\xi_{2})^{2}+(\xi_{3})^{2}}),
\label{(2.7)}
\end{eqnarray}
\begin{eqnarray}
\; & \; & {\dot A}(\xi_{1},\xi_{2},\xi_{3},t_{0})=(2 \pi)^{-3}
\int_{-\infty}^{\infty} {\rm d}x 
\int_{-\infty}^{\infty} {\rm d}y 
\int_{-\infty}^{\infty} {\rm d}z \; \Phi_{1}
{\rm e}^{-{\rm i}(\xi_{1}x+\xi_{2}y+\xi_{3}z)} \nonumber \\
&=& \sqrt{(\xi_{2})^{2}+(\xi_{3})^{2}}\Bigr[ 
Z_{1}(\xi_{1},\xi_{2},\xi_{3}){\dot J}_{{\rm i}\xi_{1}}(t_{0}\sqrt{(\xi_{2})^{2}+(\xi_{3})^{2}})
\nonumber \\
&+& Z_{2}(\xi_{1},\xi_{2},\xi_{3}){\dot Y}_{{\rm i}\xi_{1}}(t_{0}\sqrt{(\xi_{2})^{2}+(\xi_{3})^{2}}) \Bigr].
\label{(2.8)}
\end{eqnarray}
Equations (2.7) and (2.8) are a linear system of algebraic equations to be solved for $Z_{1}$ and $Z_{2}$,  
and they can be studied for various choices of Cauchy data. For example, inspired by the
simpler case of scalar wave equation in two-dimensional Minkowski spacetime, we may consider the Cauchy data
\cite{Jackson}
\begin{equation} 
\Phi_{0} \equiv {\rm e}^{-{(x^{2}+y^{2}+z^{2})\over 2L^{2}}} (\cos \gamma_{1}x)
(\cos \gamma_{2}y) (\cos \gamma_{3}z),
\label{(2.9)}
\end{equation}
\begin{equation}
\Phi_{1} \equiv 0,
\label{(2.10)}
\end{equation}
where $L$ has dimension of length. Thus, by virtue of the identity
\begin{equation}
\int_{-\infty}^{\infty}{\rm d}x \; {\rm e}^{-{\rm i}\xi x} {\rm e}^{-{x^{2}\over 2L^{2}}} (\cos \xi_{0}x)
=\sqrt{2 \pi}{L \over 2}\left[
{\rm e}^{-{L^{2}\over 2}(\xi-\xi_{0})^{2}}
+{\rm e}^{-{L^{2}\over 2}(\xi+\xi_{0})^{2}}
\right],
\label{(2.11)}
\end{equation}
we obtain from (2.7) and (2.9)
\begin{equation}
A(\xi_{1},\xi_{2},\xi_{3},t_{0})=(2 \pi)^{-{3 \over 2}} \left({L \over 2}\right)^{3}
\prod_{i=1}^{3}
\left[{\rm e}^{-{L^{2}\over 2}(\xi_{i}-\gamma_{i})^{2}}
+{\rm e}^{-{L^{2}\over 2}(\xi_{i}+\gamma_{i})^{2}}\right],
\label{(2.12)}
\end{equation}
while (2.8) and (2.10) yield
\begin{equation}
{\dot A}(\xi_{1},\xi_{2},\xi_{3},t_{0})=0.
\label{(2.13)}
\end{equation}
An interesting generalization of the Cauchy data (2.9) and (2.10) might be taken to be
\begin{equation} 
\Phi_{0} \equiv {\rm e}^{-{(x^{2}+y^{2}+z^{2})\over 2L^{2}}} (\cos \gamma_{1}x)
(\cos \gamma_{2}y) (\cos \gamma_{3}z) (\cos \gamma t_{0}),
\label{(2.14)}
\end{equation}
\begin{equation} 
\Phi_{1} \equiv \pm \gamma {\rm e}^{-{(x^{2}+y^{2}+z^{2})\over 2L^{2}}} (\cos \gamma_{1}x)
(\cos \gamma_{2}y) (\cos \gamma_{3}z)(\sin \gamma t_{0}) ,
\label{(2.15)}
\end{equation}
since it reduces to (2.9) and (2.10) at $t_{0}=0$, which is indeed the value of initial time assumed in the
Minkowski spacetime example considered in Ref. \cite{Jackson} 
(whereas in Kasner spacetime we take so far $t_{0} \not =0$ to have enough equations to
determine $Z_{1}(\xi_{1},\xi_{2},\xi_{3})$ and $Z_{2}(\xi_{1},\xi_{2},\xi_{3})$). Hereafter, to avoid cumbersome
formulas, we keep choosing the Cauchy data (2.9) and (2.10).
At this stage, Eqs. (2.7), (2.8), (2.12) and (2.13) lead to
\begin{equation}
Z_{1}(\xi_{1},\xi_{2},\xi_{3})=\left . {{\dot Y}_{{\rm i}\xi_{1}}\over 
(J_{{\rm i}\xi_{1}}{\dot Y}_{{\rm i}\xi_{1}}
-Y_{{\rm i}\xi_{1}}{\dot J}_{{\rm i}\xi_{1}})} \right |_{(t_{0}\sqrt{(\xi_{2})^{2}+(\xi_{3})^{2}})}
A(\xi_{1},\xi_{2},\xi_{3},t_{0}),
\label{(2.16)}
\end{equation}
\begin{equation}
Z_{2}(\xi_{1},\xi_{2},\xi_{3})=\left . -{{\dot J}_{{\rm i}\xi_{1}}\over 
(J_{{\rm i}\xi_{1}}{\dot Y}_{{\rm i}\xi_{1}}
-Y_{{\rm i}\xi_{1}}{\dot J}_{{\rm i}\xi_{1}})} \right |_{(t_{0}\sqrt{(\xi_{2})^{2}+(\xi_{3})^{2}})}
A(\xi_{1},\xi_{2},\xi_{3},t_{0}),
\label{(2.17)}
\end{equation}
where (2.12) should be used to express $A(\xi_{1},\xi_{2},\xi_{3},t_{0})$. The integrand of Eq. (1.4) is
therefore expressed in factorized form through Bessel functions, decaying exponentials and 
oscillating functions, but the evaluation of the integral is hard, even in this simple case.

\subsection{Representation of the solution through integration in the complex domain}

Note now that the original hyperbolic equation (1.3) is a particular case of the general form 
\cite{Garabedian}
\begin{equation}
L[u] \equiv \left[{\partial^{2}\over \partial t^{2}}-\left( \sum_{j,k=1}^{n}a^{jk}
{\partial^{2}\over \partial x^{j} \partial x^{k}}+b {\partial \over \partial t}
+\sum_{j=1}^{n}b^{j}{\partial \over \partial x^{j}}+c \right)\right]u=0.
\label{(2.18)}
\end{equation}
In the general theory, $a^{jk}$ is a symmetric tensor,
$b{\partial \over \partial t}+\sum_{j=1}^{n}b^{j}{\partial \over \partial x^{j}}$ is a
$C^{\infty}$ vector field and $c$ is a $C^{\infty}$ scalar field. In our case, 
we have $n=3,x^{1}=x,x^{2}=y,x^{3}=z$ and
\begin{equation}
a^{jk}={\rm diag}(t^{-2},1,1), \; b={1 \over t}, \; b^{j}=0, \; c=0.
\label{(2.19)}
\end{equation}
Thus, for all $t \not=0$ (as we said before, we avoid $t=0$, which is a singularity of the
Kasner coordinates), we can exploit the integral representation 
(see Appendix A) of the solution of hyperbolic equations
with variable coefficients \cite{Garabedian}, while remarking that Eq. (1.3) is also of a type
similar to other hyperbolic equations for which the mathematical literature 
(see Appendix A) has proved that the
Cauchy problem is well posed \cite{Oleinik, Menikoff}. On referring the reader to chapters $5$ and $6$
of Ref. \cite{Garabedian} for the interesting details, we simply state here the main result when 
Eqs. (2.18) and (2.19) hold.
\vskip 0.3cm
\noindent
{\bf Theorem 2.1} The solution of the scalar wave equation (1.3) with Cauchy data (2.9) and (2.10) at
$t=t_{0} \not=0$ admits the integral representation
\begin{equation}
u(t,x^{1},x^{2},x^{3})=\lim_{\partial D \to T} \int_{\partial D}
B \Bigr[u(\tau,y^{1},y^{2},y^{3}),S(\tau,y^{1},y^{2},y^{3};t,x^{1},x^{2},x^{3})\Bigr],
\label{(2.20)}
\end{equation}
where $S$ is a fundamental solution (see Appendix B) of the adjoint equation
\begin{equation}
M[S]=0,
\label{(2.21)}
\end{equation}
$M$ being the adjoint operator acting, in our case, as
\begin{equation}
M \equiv {\partial^{2}\over \partial t^{2}}-\sum_{j,k=1}^{3}a^{jk}
{\partial^{2}\over \partial x^{j} \partial x^{k}}+{{\rm d}b \over {\rm d}t}
+b {\partial \over \partial t},
\label{(2.22)}
\end{equation}
while the integrand $B[u,S]$ is the differential $3$-form
\begin{eqnarray}
\; & \; & B[u,S]=\left[bu S- \left(S {\partial u \over \partial \tau}
-u {\partial S \over \partial \tau}\right)\right] {\rm d}y^{1} \wedge {\rm d}y^{2} \wedge {\rm d}y^{3}
\nonumber \\
&+& \sum_{j=1}^{3}(-1)^{j} \sum_{k=1}^{3} a^{jk}
\left(S {\partial u \over \partial y^{k}}-u {\partial S \over \partial y^{k}}\right)
{\rm d}\tau \wedge {\rm d}y^{1} \wedge ... {\widehat {{\rm d} y^{j}}} \wedge ...
{\rm d}y^{3}.
\label{(2.23)}
\end{eqnarray}
With this notation, the hat upon ${\rm d}y^{j}$ denotes omission of integration with respect to that
particular variable, and $D$ is the region of integration viewed as a cell in the complex domain, with
boundary $\partial D$. Integration over $\partial D$ should be therefore interpreted in the sense of the
calculus of exterior differential forms. Our $D$ is a manifold defined by the conditions
\begin{equation}
{\rm Im}(\tau^{2})+\sum_{k=1}^{3} (y^{k}-x^{k})^{2} \leq \varepsilon^{2},
\label{(2.24)}
\end{equation}
\begin{equation}
{\rm Re}(\tau-t)={\rm Im}(y^{1})={\rm Im}(y^{2})={\rm Im}(y^{3})=0,
\label{(2.25)}
\end{equation}
which describe a sphere of radius $\varepsilon$ in the complex domain, centered at the real point
$(t,x^{1},x^{2},x^{3})$. Moreover, the symbolic notation $\partial D \rightarrow T$ indicates the process
of describing the boundary $\partial D$ down around the domain of dependence on the space 
where the initial data (2.9) and (2.10) are assigned (such a space is a two-dimensional plane when the Kasner
exponents $(1,0,0)$ are chosen, whereas, for more general exponents, it corresponds to a singular
surface of infinite curvature). A quite complicated evaluation of residues is involved
in Eq. (2.20), because the fundamental solution $S$ of Eq. (2.21) is singular where the 
Hadamard-Ruse-Synge world function (see Appendix B) vanishes.

\section{Bicharacteristics of the scalar wave equation in Kasner spacetime and the case
$p_{1}=p_{2}={2 \over 3},p_{3}=-{1 \over 3}$}

Equation (1.3) is just a particular case of the following general form of scalar wave equation in
Kasner spacetime:
\begin{equation}
P \phi=\left(-{\partial^{2}\over \partial t^{2}}
-{1\over t}{\partial \over \partial t}+\sum_{l=1}^{3}t^{-2p_{l}}
{\partial^{2}\over \partial {x^{l}}^{2}}\right)\phi=0.
\label{(3.1)}
\end{equation}
In light of the technical results in Appendix A, it is rather important to study Eq. (3.1)
with generic values of parameters, which is what we do now. 

The leading symbol of the wave operator $P$ is the contravariant form 
$g^{\alpha \beta}={\rm diag}(-1,t^{-2p_{1}},t^{-2p_{2}},t^{-2p_{3}})$ of the metric,
and hence the characteristic polynomial reads as
\begin{equation}
H(x,\xi)=g^{\alpha \beta}(x)\xi_{\alpha}\xi_{\beta}=-(\xi_{0})^{2}
+\sum_{k=1}^{3}t^{-2p_{k}}(\xi_{k})^{2}.
\label{(3.2)}
\end{equation}
An hypersurface $\Sigma$ is called a {\it characteristic} of $P$ if the restriction
$\left . Pu \right |_{\Sigma}$ can be expressed by using only derivatives tangential to $\Sigma$
of the restrictions $\left . u \right|_{\Sigma}$ and $\left . {\rm grad}u \right|_{\Sigma}$.
This implies that the characteristics of wave equations on $M$ are the null hypersurfaces
of $M$ \cite{Friedlander}. Hence there is, at each point of the characteristic surface 
$\Sigma$, a unique null direction that is both normal and tangential to $\Sigma$. The curves
on $\Sigma$ that are tangential to this null direction field form a congruence on $\Sigma$ and are
called the {\it bicharacteristics}.

Suppose now that $\Omega$ is a coordinate neighbourhood such that
\begin{equation}
\Omega \cap \Sigma = \left \{x: \chi(x)=0 \right \} \equiv \Sigma_{\Omega},
\label{(3.3)}
\end{equation}
where $\chi \in C^{\infty}(\Omega)$ and ${\rm grad} \chi \not =0$. If $\Sigma$ is a
characteristic, one has \cite{Friedlander,Vitagliano}
\begin{equation}
\langle {\rm grad}\chi,{\rm grad}\chi \rangle=g^{\alpha \beta}(x){\partial \chi \over \partial x^{\alpha}}
{\partial \chi \over \partial x^{\beta}}=g^{-1}({\rm d}\chi,{\rm d}\chi)=0 
\; {\rm when} \; \chi=0.
\label{(3.4)}
\end{equation}
By introducing $\chi$ as a local coordinate, this implies, for some $A \in C^{\infty}(\Omega)$, 
the equation
\begin{equation}
g^{\alpha \beta}(x)(\partial_{\alpha}\chi)(\partial_{\beta}\chi)=A \chi.
\label{(3.5)}
\end{equation}
The differential equations
\begin{equation}
{{\rm d}\over {\rm d}\tau} x^{\alpha}=g^{\alpha \beta}(x)\partial_{\beta}\chi(x)
\label{(3.6)}
\end{equation}
where $\tau$ is a parameter, have a unique smooth solution for given initial values $x^{\alpha}(0)$. Since
the gradient of $\chi$ is orthogonal to a hypersurface $\chi={\rm const}$, the bicharacteristics of $\Sigma$
are obtained by integrating Eq. (3.6) subject to the initial condition $\chi(x(0))=0$. If
the map $\tau \rightarrow x^{\alpha}(\tau)$ is an integral curve of Eq. (3.6), and
\begin{equation}
\pi_{\alpha} \equiv \partial_{\alpha}\chi(x(\tau)),
\label{(3.7)}
\end{equation}
one finds
\begin{equation}
{{\rm d}\over {\rm d}\tau}\pi_{\alpha}=(\partial_{\alpha}\partial_{\beta}\chi)
{{\rm d}\over {\rm d}\tau}x^{\beta}=(\partial_{\alpha}\partial_{\beta}\chi)
g^{\beta \gamma}\partial_{\gamma}\chi 
={1 \over 2}\partial_{\alpha} \langle {\rm grad}\chi,{\rm grad}\chi \rangle
-{1 \over 2}\pi_{\beta}\pi_{\gamma}\partial_{\alpha}g^{\beta \gamma}.
\label{(3.8)}
\end{equation}
On a bicharacteristic, one has $\chi=0 \Longrightarrow \partial_{\alpha}(A\chi)
=A \partial_{\alpha}\chi$, and hence it follows from Eq. (3.5) that the equations
\begin{equation}
{{\rm d}\over {\rm d}\tau}x^{\alpha}=g^{\alpha \beta}(x)\pi_{\beta}, \;
{{\rm d}\over {\rm d}\tau}\pi_{\alpha}=-{1 \over 2}\pi_{\beta}\pi_{\gamma}
\partial_{\alpha}g^{\beta \gamma}+{1 \over 2}A \pi_{\alpha}
\label{(3.9)}
\end{equation}
hold on a bicharacteristic. One can introduce an invertible parameter transformation by
defining a function $\tau \rightarrow \psi(\tau)$ such that \cite{Friedlander}
\begin{equation}
{{\rm d}\psi \over {\rm d}\tau}={\rm exp} \left({1 \over 2}\int_{0}^{\tau}A(x(\tau')){\rm d}\tau'\right),
\; \psi(\tau=0)=0.
\label{(3.10)}
\end{equation}
If one then defines
\begin{equation}
\xi_{\alpha} \equiv {{\rm d}\tau \over {\rm d}\psi}\pi_{\alpha}
={{\rm d}\tau \over {\rm d}\psi}\partial_{\alpha}\chi(x(\tau)),
\label{(3.11)}
\end{equation}
the equations (3.9) are transformed into the equivalent Hamilton-like set
\begin{equation}
{{\rm d}\over {\rm d}\psi}x^{\alpha}=g^{\alpha \beta}(x)\xi_{\beta}
={1 \over 2} \left . {\partial H(x,\xi) \over \partial \xi_{\alpha}} 
\right|_{ \left . H(x,\xi) \right|_{\Sigma_{\Omega}}},
\label{(3.12)}
\end{equation}
\begin{equation}
{{\rm d}\over {\rm d}\psi}\xi_{\alpha}=-{1 \over 2}\xi_{\beta}\xi_{\gamma}
\partial_{\alpha}g^{\beta \gamma}(x)
=-{1 \over 2} \left . {\partial H(x,\xi) \over \partial x^{\alpha}} 
\right|_{ \left . H(x,\xi) \right|_{\Sigma_{\Omega}}},
\label{(3.13)}
\end{equation}
where ${1 \over 2}H(x,\xi)$ plays the role of Hamiltonian function. Interestingly,
Eqs. (3.12) and (3.13) are the equations of null geodesics, in canonical form. The null
nature of such geodesics follows immediately from the definition (3.11) and Eq. (3.5), i.e.
\begin{equation}
\left . H(x,\xi) \right|_{\Sigma_{\Omega}}
=\Bigr[g^{\alpha \beta}(x)\xi_{\alpha}\xi_{\beta}\Bigr]_{\Sigma_{\Omega}}
=\left[\left({{\rm d}\tau \over {\rm d}\psi}\right)^{2}g^{\alpha \beta}
(\partial_{\alpha}\chi)(\partial_{\beta}\chi)\right]_{\Sigma_{\Omega}}
=\left[\left({{\rm d}\tau \over {\rm d}\psi}\right)^{2} A \chi \right]_{\Sigma_{\Omega}}=0.
\label{(3.14)}
\end{equation}
Thus, the bicharacteristics are null geodesics.

A characteristic contains a congruence of bicharacteristics, the latter being the null geodesics
of spacetime $(M,g)$. The bicharacteristics of $P$ can also be viewed as the projections on the
manifold $M$ of the curves in the cotangent bundle $T^{*}M$, called bicharacteristic strips, that
are the integral curves of the Hamiltonian system of ordinary differential equations (3.12) 
and (3.13). The equation to which Eq. (3.5) reduces when $\chi=0$ 
(which occurs on characteristics and bicharacteristics) is the well known 
{\it eikonal equation} (cf. Ref. \cite{Choquet1968}). If the parameter $\psi$ is set equal to $2 \zeta$,
there is complete formal analogy between Eqs. (3.12), (3.13) and a set of Hamilton equations. 
As far as wave propagation is concerned,
the {\it wave fronts} of our wave equation (3.1) are characteristic surfaces (satisfying therefore
Eqs. (3.3) and (3.5)) and propagate along bicharacteristics \cite{Garabedian, LeviCivita}.
With our Kasner metric and $\psi=2 \zeta$, 
the Eqs. (3.12) and (3.13) for bicharacteristics,
bearing in mind that $x^{0}=t$ in $c=1$ units, take the form
(there is no summation over $i$ in Eq. (3.16) below) 
\begin{equation}
{{\rm d}t \over {\rm d}\zeta}=-2 \xi_{0}(\zeta),
\label{(3.15)}
\end{equation}
\begin{equation}
{{\rm d}x^{i}\over {\rm d}\zeta}=2 \xi_{i}(t(\zeta))^{-2p_{i}},
\label{(3.16)}
\end{equation}
\begin{equation}
{{\rm d}\xi_{0}\over {\rm d}\zeta}=2\sum_{k=1}^{3}p_{k}(\xi_{k})^{2}
(t(\zeta))^{-2p_{k}-1},
\label{(3.17)}
\end{equation}
\begin{equation}
{{\rm d}\xi_{i}\over {\rm d}\zeta}=0.
\label{(3.18)}
\end{equation}
Equation (3.18) is solved by
\begin{equation}
\xi_{i}={\tilde \xi}_{i} \; {\rm constant} \; \forall i=1,2,3.
\label{(3.19)}
\end{equation}
Further differentiation with respect to $\zeta$ of Eq. (3.15) leads therefore,
by virtue of (3.17), to the equivalent decoupled system given by
\begin{equation}
{{\rm d}^{2}t \over {\rm d}\zeta^{2}}=-4 \sum_{k=1}^{3}p_{k}
({\tilde \xi}_{k})^{2}(t(\zeta))^{-2p_{k}-1}
\label{(3.20)}
\end{equation}
together with Eqs. (3.16) and (3.17). Upon defining 
$A_{k} \equiv -4 p_{k}({\tilde \xi}_{k})^{2}$ we find, more explicitly, the following
nonlinear equation for $t(\zeta)$:
\begin{equation}
{{\rm d}^{2}t \over {\rm d}\zeta^{2}}={A_{1}\over t^{2p_{1}+1}}
+{A_{2}\over t^{2p_{2}+1}}+{A_{3}\over t^{2p_{3}+1}}=f(t),
\label{(3.21)}
\end{equation}
supplemented by the initial conditions 
\begin{equation}
t'(\zeta=0)=-2\xi_{0}(0), \; t(\zeta=0)=t_{0}.
\label{(3.22)}
\end{equation}
Since the right-hand side of Eq. (3.21) is independent of $\zeta$ we are dealing with an autonomous 
differential equation, which can be solved by separation of variables. For this purpose, setting
$t'={{\rm d}t \over {\rm d}\zeta}=Q(t)$ one finds
\begin{equation}
t''={{\rm d}Q \over {\rm d}t}{{\rm d}t \over {\rm d}\zeta}
=Q{{\rm d}Q \over {\rm d}t}=f(t) \Longrightarrow Q {\rm d}Q=f(t){\rm d}t,
\label{(3.23)}
\end{equation}
which implies, denoting by $\gamma$ an integration constant,
\begin{equation}
Q^{2}=\gamma +2 \int f(t){\rm d}t,
\label{(3.24)}
\end{equation}
and hence
\begin{equation}
Q= \pm \sqrt{\gamma + 2 \int f(t){\rm d}t}={{\rm d}t \over {\rm d}\zeta},
\label{(3.25)}
\end{equation}
where a further integration yields
\begin{equation}
\int {{\rm d}t \over \sqrt{\gamma +2 \int f(t) {\rm d}t}}= \pm \zeta + \kappa,
\label{(3.26)}
\end{equation}
for some constant $\kappa$. In our case, the integral of $f$ is such that
\begin{equation}
2 \int f(t){\rm d}t= 2 \sum_{k=1}^{3} -{A_{k}\over 2p_{k}} t^{-2p_{k}}
=\sum_{k=1}^{3} 4 ({\tilde \xi}_{k})^{2}t^{-2p_{k}},
\label{(3.27)}
\end{equation}
so that Eq. (3.26) for the geodesic parameter $\zeta$ in terms of the time variable $t$ reads as
\begin{equation}
\pm \zeta(t)=\int {{\rm d}t \over \sqrt{\gamma+ \sum_{k=1}^{3} 4({\tilde \xi}_{k})^{2}
t^{-2p_{k}}}}- \kappa.
\label{(3.28)}
\end{equation}
As far as we can see, such an integral cannot be evaluated explicitly for generic values of
Kasner parameters, but in the particular case
\begin{equation}
p_{1}=p_{2}={2 \over 3}, \; p_{3}=-{1 \over 3},
\label{(3.29)}
\end{equation}
which is technically harder than the $(1,0,0)$ choice studied in Ref. \cite{Nariai}, we have found an
explicit formula in terms of elliptic integrals. To be self-contained, we recall some basic
definitions, as follows.
\vskip 0.3cm
\noindent
(i) If $\Omega \in \left]-{\pi \over 2},{\pi \over 2} \right[$, the elliptic integral of the first
kind, here denoted by $E_{I}[\Omega,m]$, is defined by
\begin{equation}
E_{I}[\Omega,m] \equiv \int_{0}^{\Omega}(1-m \sin^{2}\theta)^{-{1 \over 2}}{\rm d}\theta.
\label{(3.30)}
\end{equation}
(ii) If $\Omega \in \left]-{\pi \over 2},{\pi \over 2} \right[$, the elliptic integral of the
second kind, $E_{II}[\Omega,m]$, reads as
\begin{equation}
E_{II}[\Omega,m] \equiv \int_{0}^{\Omega}(1-m \sin^{2}\theta)^{{1 \over 2}}{\rm d}\theta.
\label{(3.31)}
\end{equation}
(iii) If $\Omega$ lies the same open interval as in (i) and (ii), the incomplete elliptic integral
of the third kind is given by
\begin{equation}
E_{III}[n
,\Omega,m] \equiv \int_{0}^{\Omega}(1-n \sin^{2}\theta)^{-1}
(1-m \sin^{2}\theta)^{-{1 \over 2}}{\rm d}\theta.
\label{(3.32)}
\end{equation}

Moreover, in order to express our results, we have to consider the three roots 
$r_{1},r_{2},r_{3}$ of the algebraic equation
\begin{equation}
x^{3}+{\gamma \over 4 ({\tilde \xi}_{3})^{2}}x^{2}
+{\Bigr[({\tilde \xi}_{1})^{2}+({\tilde \xi}_{2})^{2}\Bigr]\over
({\tilde \xi}_{3})^{2}}=0.
\label{(3.33)}
\end{equation}
If the three roots are all real, we follow the convention according to which 
$r_{1}< r_{2} < r_{3}$. If $r_{1}$ is the real root while the remaining two roots are complex
conjugate, we agree that $r_{2}$ and $r_{3}$ are such that ${\rm Im}r_{2} < {\rm Im}r_{3}$. 
With this understanding, and defining
\begin{equation}
f_{k,1}(r) \equiv {r_{k} \over (r_{k}-r_{1})}, k=2,3,
\label{(3.34)}
\end{equation}
\begin{equation}
f(r,t) \equiv \sqrt{{t^{2/3}\over f_{3,1}(r)(t^{2/3}-r_{1})}},
\label{(3.35)}
\end{equation}
\begin{eqnarray}
H(\xi,r,t)& \equiv & 4(({\tilde \xi}_{1})^{2}+({\tilde \xi}_{2})^{2})
\sqrt{{r_{1}(t^{2/3}-r_{2}) \over r_{2}(t^{2/3}-r_{1})}}
\nonumber \\
& \times & \left \{ 4({\tilde \xi}_{3})^{2} \left({r_{3} \over r_{1}}-1 \right)
E_{II}[{\rm arcsin}f(r,t),f_{3,1}(r)/f_{2,1}(r)] \right .
\nonumber \\
& + & {1 \over r_{2}}\Bigr[(\gamma +4 ({\tilde \xi}_{3})^{2}r_{2})
E_{I}[{\rm arcsin}f(r,t),f_{3,1}(r)/f_{2,1}(r)]
\nonumber \\
& - & \left . \gamma E_{III}[f_{3,1}(r),{\rm arcsin}f(r,t),f_{3,1}(r)/f_{2,1}(r)]\Bigr]\right \},
\label{(3.36)}
\end{eqnarray}
the integral (3.28) reads as
\begin{eqnarray}
\pm \zeta(t)&=& {1 \over 2}\left({4(({\tilde \xi}_{1})^{2}+({\tilde \xi}_{2})^{2})
+\gamma t^{4 \over 3}+4 ({\tilde \xi}_{3})^{2}t^{2} \over t^{2/3}}\right)^{-1/2}
\biggr[3(t^{2/3}-r_{3}) 
\nonumber \\
& \times & \left . \left(t^{2/3}-r_{2}
-{H(\xi,r,t)(t^{2/3}-r_{1})\over 16 ({\tilde \xi}_{3})^{4}(r_{3})^{2}
\sqrt{{r_{1}(t^{2/3}-r_{3}) \over r_{3}f_{3,1}(r)}}}\right)\right].
\label{(3.37)}
\end{eqnarray}

Furthermore, since Eq. (3.37) cannot be explicitly inverted to express $t=t(\zeta)$, it is more
convenient to use the identities
\begin{equation}
{{\rm d}x^{i}\over {\rm d}t}={{\rm d}x^{i}\over {\rm d}\zeta} {{\rm d}\zeta \over {\rm d}t}, \;
{{\rm d}\xi_{0} \over {\rm d}t}={{\rm d}\xi_{0}\over {\rm d}\zeta} 
{{\rm d}\zeta \over {\rm d}t},
\label{(3.38)}
\end{equation}
so that the remaining components of null geodesic equations are expressed, from (3.26) and (3.38),
in the form
\begin{equation}
\int {\rm d}x^{i}= \pm 2 {\tilde \xi}_{i}
\int {t^{-2 p_{i}} \over \sqrt{\gamma + \sum_{k=1}^{3}4 ({\tilde \xi}_{k})^{2}t^{-2p_{k}}}}
{\rm d}t,
\label{(3.39)}
\end{equation}
\begin{equation}
\int {\rm d}\xi_{0}= \pm 2 
\int {\sum_{k=1}^{3}p_{k}({\tilde \xi}_{k})^{2}t^{-2p_{k}-1} 
\over \sqrt{\gamma + \sum_{k=1}^{3} 4({\tilde \xi}_{k})^{2} t^{-2p_{k}}}}
{\rm d}t.
\label{(3.40)}
\end{equation}
With the choice (3.29) of Kasner parameters, also (3.39) can be expressed in terms
of elliptic integrals, while (3.40) takes the remarkably simple form
\begin{equation}
\xi_{0}(t)= \mp {1 \over 2} \sqrt{{4 \Bigr[({\tilde \xi}_{1})^{2}
+({\tilde \xi}_{2})^{2}\Bigr]+\gamma t^{4 \over 3}
+4({\tilde \xi}_{3})^{2}t^{2} \over t^{{4 \over 3}}}}.
\label{(3.41)}
\end{equation}
In the formula (3.39), when $i=1,2$, we find
\begin{equation}
{\int {\rm d}x^{1} \over 2 {\tilde \xi}_{1}}
={\int {\rm d}x^{2} \over 2 {\tilde \xi}_{2}}=\pm I(t),
\label{(3.42)}
\end{equation}
where
\begin{equation}
I(t)= \mp {3 \over 2}{t^{-2/3}\over \xi_{0}}
E_{I}[{\rm arcsin}f(r,t),f_{3,1}(r)/f_{2,1}(r)]
\sqrt{(t^{2/3}-r_{1})(t^{2/3}-r_{2})(t^{2/3}-r_{3})
{f_{3,1}(r)\over r_{2}r_{3}}}, 
\label{(3.43)}
\end{equation}
while, if $i=3$, we find
\begin{equation}
{\int {\rm d}x^{3} \over 2 {\tilde \xi}_{3}}=\pm I_{3}(t),
\label{(3.44)}
\end{equation}
where, upon defining
\begin{equation}
A^{2} \equiv 4 ({\tilde \xi}_{1})^{2}, \;
B^{2} \equiv 4 ({\tilde \xi}_{2})^{2}, \;
C^{2} \equiv 4 ({\tilde \xi}_{3})^{2}, \;
\rho(t) \equiv \sqrt{A^{2}+B^{2}+\gamma t^{4/3}+C^{2}t^{2}},
\label{(3.45)}
\end{equation}
we write
\begin{eqnarray}
I_{3}(t) & \equiv & {3 \over 4 C^{2}} \sqrt{\rho(t)}t^{1/3}
-{3 \over 8 C^{2}} \biggr \{ - \biggr[ \biggr(
2(A^{2}+B^{2})r_{3}(t^{2/3}-r_{1})^{2}
\nonumber \\ 
& \times & {\sqrt{(t^{2/3}-r_{2})(t^{2/3}-r_{3})} \over
(t^{2/3}-r_{1})^{3/2} (r_{3}-r_{1})\sqrt{\rho(t)r_{2}r_{3}}}
E_{I}[{\rm arcsin}f(r,t),f_{3,1}(r)/f_{2,1}(r)]\biggr)
\nonumber \\
&+& 3 \gamma \biggr((t^{2/3}-r_{2})(t^{2/3}-r_{3})t^{2/3}
\nonumber \\
&+& r_{3} (t^{2/3}-r_{1})^{2}
{\sqrt{(t^{2/3}-r_{2})(t^{2/3}-r_{3})}r_{1} \over
(t^{2/3}-r_{1})^{3/2}\sqrt{\rho(t)r_{2}r_{3}}}
\nonumber \\
& \times & \biggr({(r_{1}+r_{3})\over (r_{1}-r_{3})}
E_{I}[{\rm arcsin}f(r,t),f_{3,1}(r)/f_{2,1}(r)]
+{r_{2}\over r_{1}}
E_{II}[{\rm arcsin}f(r,t),f_{3,1}(r)/f_{2,1}(r)]
\nonumber \\
&+& {(r_{1}+r_{2}+r_{3})\over (r_{3}-r_{1})}
E_{III}[f_{3,1}(r),{\rm arcsin}f(r,t),f_{3,1}(r)/f_{2,1}(r)]\biggr)
\biggr)\biggr] \biggr \}.
\label{(3.46)}
\end{eqnarray}
Our original calculation of bicharacteristics in Kasner spacetime is therefore completed.

\section{Parametrix for the Cauchy problem through Fourier-Maslov integral operators: the general case}

Since we are studying a wave equation, we may expect that the solution formula involves amplitude
and phase functions, as well as the Cauchy data (here, unlike Sec. II, we exploit techniques that do
not need to avoid $t=0$)
\begin{equation}
\phi(t=0,x) \equiv u_{0}(x),
\label{(4.1)}
\end{equation}
\begin{equation}
{\partial \phi \over \partial t}(t=0,x) \equiv u_{1}(x),
\label{(4.2)}
\end{equation}
which are again assumed to be Fourier transformable. However, the variable nature of the coefficients
demands for a nontrivial generalization of the integral representation (1.4). This is indeed available,
since a theorem guarantees that the solution of the Cauchy problem (3.1), (4.1) 
and (4.2) can be expressed in the form \cite{Treves}
\begin{equation}
\phi(x,t)=\sum_{j=0}^{1}E_{j}(t)u_{j}(x),
\label{(4.3)}
\end{equation}
where, on denoting by ${\hat u}_{j}$ the Fourier transform of the Cauchy data, the operators 
$E_{j}(t)$ act according to (hereafter, $(x) \equiv (x^{1},x^{2},x^{3})$, with covariable
$(\xi) \equiv (\xi_{1},\xi_{2},\xi_{3})$)
\begin{equation}
E_{j}(t)u_{j}(x)= \sum_{k=1}^{2}(2\pi)^{-3}
\int {\rm e}^{{\rm i}\varphi_{k}(x,t,\xi)}\alpha_{jk}(x,t,\xi){\hat u}_{j}(\xi){\rm d}^{3}\xi
+R_{j}(t)u_{j}(x),
\label{(4.4)}
\end{equation}
where the $\varphi_{k}$ are real-valued phase functions which satisfy the initial condition
\begin{equation}
\varphi_{k}(t=0,x,\xi)=x \cdot \xi = \sum_{s=1}^{3}x^{s}\xi_{s},
\label{(4.5)}
\end{equation}
and $R_{j}(t)$ is a regularizing operator which smoothes out the singularities acted upon by it
\cite{Treves}. In other words, the Cauchy problem is here solved by a pair of Fourier-Maslov
integral operators of the form (4.4), and such a construction (leaving aside, for the moment,
its global version, which can be built as shown in Chapter VII of Ref. \cite{Treves})
generalizes the monochromatic plane waves for the d'Alembert operator from Minkowski spacetime
to Kasner spacetime. Strictly, we are dealing with the approximate Green function for the wave
equation, called the parametrix. In our case, since we know a priori that (4.3) and (4.4) yield
an exact solution of the Cauchy problem, 
we can insert them into Eq. (3.1), finding that, for all $j=0,1$,
\begin{equation}
P[E_{j}(t)u_{j}(x)] \sim \sum_{k=1}^{2} (2\pi)^{-3}
\int P[{\rm e}^{{\rm i}\varphi_{k}}\alpha_{jk}] {\hat u}_{j}(\xi){\rm d}^{3}\xi,
\label{(4.6)}
\end{equation}
where $P R_{j}(t)u_{j}(x)$ can be neglected with respect to the integral on the right-hand side of
Eq. (4.4), because $R_{j}(t)$ is a regularizing operator. Next, we find from Eq. (3.1) that
\begin{equation}
P[{\rm e}^{{\rm i}\varphi_{k}}\alpha_{jk}]={\rm e}^{{\rm i}\varphi_{k}}({\rm i}A_{jk}+B_{jk}),
\label{(4.7)}
\end{equation}
where
\begin{equation}
A_{jk} \equiv {\partial^{2}\varphi_{k}\over \partial t^{2}}\alpha_{jk}
+2 {\partial \varphi_{k}\over \partial t}{\partial \alpha_{jk}\over \partial t}
+{1 \over t}{\partial \varphi_{k}\over \partial t}\alpha_{jk}
-\sum_{l=1}^{3}t^{-2p_{l}}
\left({\partial^{2}\varphi_{k}\over \partial x_{l}^{2}}\alpha_{jk}
+2 {\partial \varphi_{k}\over \partial x_{l}}
{\partial \alpha_{jk}\over \partial x_{l}} \right),
\label{(4.8)}
\end{equation}
\begin{equation}
B_{jk} \equiv {\partial^{2}\alpha_{jk}\over \partial t^{2}}
-\left({\partial \varphi_{k}\over \partial t}\right)^{2}\alpha_{jk}
+{1 \over t}{\partial \alpha_{jk}\over \partial t}
-\sum_{l=1}^{3}t^{-2p_{l}}
\left({\partial^{2}\alpha_{jk}\over \partial x_{l}^{2}}
-\left({\partial \varphi_{k}\over \partial x_{l}}\right)^{2}
\alpha_{jk}\right).
\label{(4.9)}
\end{equation}
If the phase functions $\varphi_{k}$ are real-valued, since
the exponentials ${\rm e}^{{\rm i}\varphi_{k}}$  
can be taken to be linearly independent, we can fulfill Eq. (3.1),
up to the negligible contributions resulting from $P R_{j}(t)u_{j}(x)$, by setting to zero
in the integrand (4.6) both $A_{jk}$ and $B_{jk}$. This leads to a coupled system of partial
differential equations. Our Cauchy problem (3.1), (4.1)
and (4.2) is therefore equivalent to solving the equations
\begin{equation}
A_{jk}=0, \; B_{jk}=0.
\label{(4.10)}
\end{equation}

Equation (4.10) is the {\it dispersion relation} for the scalar wave equation in Kasner spacetime.
Such a dispersion relation takes a neater geometric form upon bearing in mind the form (3.1) of the 
wave (or d'Alembert) operator $P= \Box$ in Kasner coordinates, i.e.
\begin{equation}
A_{jk}=0 \Longrightarrow \Bigr[-\alpha_{jk}(\Box \varphi_{k})
-2g^{\beta \gamma}(\varphi_{k})_{,\beta}(\alpha_{jk})_{,\gamma}\Bigr]=0,
\label{(4.11)}
\end{equation}
\begin{equation}
B_{jk}=0 \Longrightarrow \Bigr[-\Box+ g^{\beta \gamma}(\varphi_{k})_{,\beta}
(\varphi_{k})_{,\gamma}\Bigr]\alpha_{jk}=0.
\label{(4.12)}
\end{equation}
Let us bear in mind that the indices $j$ and $k$ are not tensorial, but they merely count the number of
functions contributing to the Fourier-Maslov integral operator (4.4). We can therefore exploit the
four-dimensional concept of gradient of a function \cite{Choquet1968} as the four-dimensional
covariant vector defined by the differential of the function, i.e.
\begin{equation}
{\rm d}f={\partial f \over \partial x^{\alpha}}{\rm d}x^{\alpha}
=f_{,\alpha}{\rm d}x^{\alpha}=(\nabla_{\alpha}f){\rm d}x^{\alpha}
=({\rm grad}f)_{\alpha}{\rm d}x^{\alpha},
\label{(4.13)}
\end{equation}
where $\nabla$ is the Levi-Civita connection on four-dimensional spacetime, and we exploit the identity
$f_{,\alpha}=\nabla_{\alpha}f, \; \forall f \in C^{\infty}(M)$. The consideration of $\nabla_{\alpha}f$
is not mandatory at this stage, but it will be helpful in a moment, when we write (see below) in tensor
language the equations expressing the dispersion relation. 

We arrive therefore, upon multiplying
Eq. (4.11) by $\alpha_{jk}$, while dividing Eq. (4.12) by $\alpha_{jk}$, at the following geometric
form of dispersion relation in Kasner spacetime (with our notation we actually write it in the
same way in any Lorentzian spacetime):
\begin{equation}
g^{\beta \gamma}\nabla_{\beta}\Bigr[(\alpha_{jk})^{2}\nabla_{\gamma}\varphi_{k}\Bigr]
={\rm div}\Bigr[(\alpha_{jk})^{2}{\rm grad}\varphi_{k}\Bigr]=0,
\label{(4.14)}
\end{equation}
\begin{equation}
g^{\beta \gamma}(\nabla_{\beta}\varphi_{k})(\nabla_{\gamma}\varphi_{k})
=\langle {\rm grad}\varphi_{k},{\rm grad}\varphi_{k} \rangle ={(\Box \alpha_{jk})\over \alpha_{jk}},
\label{(4.15)}
\end{equation}
where the four-dimensional divergence operator acts according to\footnote{In particular, when
$F$ is a gradient, one gets therefore the wave operator on scalars, i.e.
$$
\Box \equiv {\rm div} \; {\rm grad}=g^{\alpha \beta}\nabla_{\alpha}\nabla_{\beta}.
$$}
\begin{equation}
{\rm div}F=\nabla^{\beta}F_{\beta}=g^{\alpha \beta}\nabla_{\alpha}F_{\beta}.
\label{(4.16)}
\end{equation}

Note that, if the ratio ${(\Box \alpha_{jk})\over \alpha_{jk}}$ is much smaller than a suitable
parameter having dimension ${\rm length}^{-2}$, Eq. (4.15) reduces to the eikonal equation and
hence the phase function reduces to the Hadamard-Ruse-Synge world function of Appendices B and C.
This property makes contact with the asymptotic expansion presented in Appendix B. However, it is
possible to devise a strategy to solve exactly Eqs. (4.14) and (4.15). For this purpose we remark
that, upon defining the covariant vector
\begin{equation}
\psi_{\gamma} \equiv (\alpha_{jk})^{2} \nabla_{\gamma}\varphi_{k},
\label{(4.17)}
\end{equation}
Eq. (4.14) is equivalent to solving the first-order partial differential equation expressing the
vanishing divergence condition for $\psi_{\gamma}$, i.e.
\begin{equation}
\nabla^{\gamma}\psi_{\gamma}={\rm div}\psi=0.
\label{(4.18)}
\end{equation}
Of course, this equation is not enough to determine the four components of $\psi_{\gamma}$,
but there are cases where further progress can be made (see below).
After doing that, we can express the (covariant) derivative of the phase function from the
definition (4.17), i.e.
\begin{equation}
\nabla_{\gamma}\varphi_{k}=\partial_{\gamma}\varphi_{k}=(\alpha_{jk})^{-2}\psi_{\gamma},
\label{(4.19)}
\end{equation}
and the insertion of Eq. (4.19) into Eq. (4.15) yields
\begin{equation}
(\alpha_{jk})^{3}\Box \alpha_{jk}=g(\psi,\psi)=g^{\beta \gamma}\psi_{\beta}\psi_{\gamma}
=\psi_{\gamma}\psi^{\gamma}.
\label{(4.20)}
\end{equation}
Interestingly, this is a tensorial generalization of a famous nonlinear ordinary differential
equation, i.e. the Ermakov-Pinney equation \cite{Ermakov,Pinney,Lewis}
\begin{equation}
y''+py=q y^{-3}.
\label{(4.21)}
\end{equation}
If $y''$ is replaced by $\Box y$, $p$ is set to zero and $q$ is promoted to a function of spacetime
location, Eq. (4.21) is mapped into Eq. (4.20).
After solving this nonlinear equation for $\alpha_{jk}=\alpha_{jk}[g(\psi,\psi)]$, the task remains
of finding the phase function $\varphi_{k}$ by writing and solving the four components of
Eq. (4.19). To sum up, we have proved the following original result.
\vskip 0.3cm
\noindent
{\bf Theorem 4.1} For any Lorentzian spacetime manifold $(M,g)$, the amplitude functions
$\alpha_{jk} \in C^{2}(T^{*}M)$ and phase functions $\varphi_{k} \in C^{1}(T^{*}M)$ in the
parametrix (4.4) for the scalar wave equation can be obtained by solving, first, the linear condition
(4.18) of vanishing divergence for a covariant vector $\psi_{\gamma}$. All nonlinearities of the
coupled system are then mapped into solving the nonlinear equation (4.20) for the amplitude function
$\alpha_{jk}$. Eventually, the phase function $\varphi_{k}$ is found by solving the first-order 
linear equation (4.19).

In Kasner spacetime, Eq. (4.18) takes indeed the form 
\begin{equation}
\left({\partial \over \partial t}+{1 \over t}\right)\psi_{0}
=\sum_{l=1}^{3}t^{-2p_{l}}{\partial \psi_{l}\over \partial x^{l}}.
\label{(4.22)}
\end{equation}
This suggests considering ${\tilde \psi}_{0}$ and ${\tilde \psi}_{l}$ such that
\begin{equation}
\psi_{0}={1 \over t}{\tilde \psi}_{0}, \;
\psi_{l}=t^{2p_{l}-1}{\tilde \psi}_{l} \; \forall l=1,2,3,
\label{(4.23)}
\end{equation}
so that Eq. (4.22) leads to the equation
\begin{equation}
{\partial {\tilde \psi}_{0} \over \partial t}
=\sum_{l=1}^{3}{\partial {\tilde \psi}_{l} \over \partial x^{l}}.
\label{(4.24)}
\end{equation}
This is precisely the vanishing divergence condition satisfied by retarded potentials
in Minkowski spacetime in the coordinates $(t,x,y,z)$. Their integral representation
is well known to be of the form (recall that we work in $c=1$ units)
\begin{equation}
{\tilde \psi}_{0}= \int \int \int {\rho (t-r(x,y,z;x',y',z')) \over r(x,y,z;x',y',z')}
{\rm d}x' \; {\rm d}y' \; {\rm d}z',
\label{(4.25)}
\end{equation}
\begin{equation}
{\tilde \psi}_{l}= \int \int \int {s_{l} (t-r(x,y,z;x',y',z')) \over r(x,y,z;x',y',z')}
{\rm d}x' \; {\rm d}y' \; {\rm d}z',
\label{(4.26)}
\end{equation}
where
\begin{equation}
r \equiv \sqrt{(x-x')^{2}+(y-y')^{2}+(z-z')^{2}},
\label{(4.27)}
\end{equation}
and hence Eqs. (4.23), (4.25) and (4.26) solve completely the problem of finding the auxiliary
covariant vector $\psi_{\gamma}$ in Kasner spacetime.

We should now solve Eq. (4.20) for $\alpha_{jk}$. The reader might wonder what has been gained by
turning the task of solving the scalar wave equation into the task of solving Eq. (4.20). In this
equation, we can first get rid of the part linear in ${\partial \over \partial t}$ in the 
$\Box$ operator by setting
\begin{equation}
\alpha_{jk}={1 \over \sqrt{t}}{\tilde \alpha}_{jk},
\label{(4.28)}
\end{equation}
which leads to
\begin{equation}
({\tilde \alpha}_{jk})^{3}\left[-{\partial^{2}\over \partial t^{2}}-{1 \over 4 t^{2}}
+\sum_{l=1}^{3}t^{-2p_{l}}{\partial^{2}\over \partial {x^{l}}^{2}} \right]
{\tilde \alpha}_{jk}=t^{2}\psi_{\gamma}\psi^{\gamma}.
\label{(4.29)}
\end{equation}
Next, we can get rid of powers of ${\tilde \alpha}_{jk}$ by setting 
${\tilde \alpha}_{jk} \equiv (f_{jk})^{\beta}$, which yields
\begin{equation}
({\tilde \alpha}_{jk})^{3}{\partial^{2}{\tilde \alpha}_{jk}\over \partial t^{2}}
=\beta (f_{jk})^{4 \beta -1}{\partial^{2}f_{jk}\over \partial t^{2}}
+\beta (\beta-1)(f_{jk})^{4 \beta -2}\left({\partial f_{jk}\over \partial t}\right)^{2},
\label{(4.30)}
\end{equation}
and the same formula holds with $t$ replaced by $x^{l}$. Thus, upon choosing $\beta={1 \over 4}$, 
we obtain eventually the amplitude functions by solving the following nonlinear equation 
for $f_{jk}$:
\begin{eqnarray}
\; & \; & L f_{jk} \equiv \left({\partial^{2}\over \partial t^{2}}+{1 \over t^{2}}
-\sum_{l=1}^{3}t^{-2p_{l}}{\partial^{2}\over \partial {x^{l}}^{2}}\right)f_{jk}
\nonumber \\
&=& {3 \over 4}(f_{jk})^{-1}\left[\left({\partial f_{jk}\over \partial t}\right)^{2}
-\sum_{l=1}^{3}t^{-2p_{l}}\left({\partial f_{jk}\over \partial x^{l}}\right)^{2}\right]
+4t^{2}\Bigr[(\psi_{0})^{2}-\sum_{l=1}^{3}t^{-2p_{l}}(\psi_{l})^{2}\Bigr].
\label{(4.31)}
\end{eqnarray}

The form (4.31) of the equation for $f_{jk}$ makes it possible to apply the powerful Adomian
method \cite{Adomian} for the solution of nonlinear partial differential equations. For this 
purpose, inspired by Ref. \cite{Adomian}, we define the four linear operators occurring
in $L$, i.e.
\begin{equation}
L_{t} \equiv {\partial^{2}\over \partial t^{2}}, \;
L_{x} \equiv t^{-2p_{1}}{\partial^{2} \over \partial x^{2}}, \;
L_{y} \equiv t^{-2p_{2}}{\partial^{2} \over \partial y^{2}}, \;
L_{z} \equiv t^{-2p_{3}}{\partial^{2} \over \partial z^{2}}, 
\label{(4.32)}
\end{equation}
the remainder (i.e., lower order part) of the linear operator $L$, i.e.
\begin{equation}
R \equiv {1 \over t^{2}},
\label{(4.33)}
\end{equation}
the nonlinear term (hereafter we omit the subscripts $j,k$ for simplicity of notation)
\begin{equation}
Nf \equiv {3 \over 4}f \left[\sum_{l=1}^{3}t^{-2p_{l}}
\left({1 \over f}{\partial f \over \partial x^{l}}\right)^{2}
-\left({1 \over f}{\partial f \over \partial t}\right)^{2}\right],
\label{(4.34)}
\end{equation}
while the part of the right-hand side which is independent of $f$ is here denoted by $\eta$, i.e.
\begin{equation}
\eta \equiv 4t^{2}\left[(\psi_{0})^{2}-\sum_{l=1}^{3}t^{-2p_{l}}(\psi_{l})^{2}\right].
\label{(4.35)}
\end{equation}
Hence the nonlinear equation (4.31) can be re-expressed in the form
\begin{equation}
(L_{t}-L_{x}-L_{y}-L_{z})f+(R+N)f=\eta.
\label{(4.36)}
\end{equation}
The idea is now to apply the inverse of $L_{t}$, or $L_{x}$, or $L_{y}$, or $L_{z}$ to this
equation, which, upon bearing in mind the identities \cite{Adomian}
\begin{equation}
L_{t}^{-1}L_{t}f=f-\alpha_{1}-\alpha_{2}t,
\label{(4.37)}
\end{equation}
\begin{equation}
L_{x}^{-1}L_{x}f=f-\alpha_{3}-\alpha_{4}x,
\label{(4.38)}
\end{equation}
\begin{equation}
L_{y}^{-1}L_{y}f=f-\alpha_{5}-\alpha_{6}y,
\label{(4.39)}
\end{equation}
\begin{equation}
L_{z}^{-1}L_{z}f=f-\alpha_{7}-\alpha_{8}z,
\label{(4.40)}
\end{equation}
the $\alpha$'s being constants fixed by the initial and boundary conditions, leads to the following
four equations:
\begin{equation}
f=\alpha_{1}+\alpha_{2}t+L_{t}^{-1}(L_{x}+L_{y}+L_{z})f-L_{t}^{-1}Rf
-L_{t}^{-1}Nf+L_{t}^{-1}\eta,
\label{(4.41)}
\end{equation}
\begin{equation}
f=\alpha_{3}+\alpha_{4}x+L_{x}^{-1}(L_{t}-L_{y}-L_{z})f+L_{x}^{-1}Rf
+L_{x}^{-1}Nf-L_{x}^{-1}\eta,
\label{(4.42)}
\end{equation}
\begin{equation}
f=\alpha_{5}+\alpha_{6}y+L_{y}^{-1}(L_{t}-L_{z}-L_{x})f+L_{y}^{-1}Rf
+L_{y}^{-1}Nf-L_{y}^{-1}\eta,
\label{(4.43)}
\end{equation}
\begin{equation}
f=\alpha_{7}+\alpha_{8}z+L_{z}^{-1}(L_{t}-L_{x}-L_{y})f+L_{z}^{-1}Rf
+L_{z}^{-1}Nf-L_{z}^{-1}\eta.
\label{(4.44)}
\end{equation}
Now we add these four equations, and upon defining
\begin{eqnarray}
K & \equiv & {1 \over 4}\Bigr[L_{t}^{-1}(L_{x}+L_{y}+L_{z}-R)+L_{x}^{-1}(L_{t}-L_{y}-L_{z}+R)
\nonumber \\
&+& L_{y}^{-1}(L_{t}-L_{z}-L_{x}+R)+L_{z}^{-1}(L_{t}-L_{x}-L_{y}+R)\Bigr],
\label{(4.45)}
\end{eqnarray}
\begin{equation}
G \equiv {1 \over 4}(L_{t}^{-1}-L_{x}^{-1}-L_{y}^{-1}-L_{z}^{-1}),
\label{(4.46)}
\end{equation}
\begin{equation}
f_{0} \equiv {1 \over 4}\Bigr[\alpha_{1}+\alpha_{2}t+ \alpha_{3}+\alpha_{4}x
+\alpha_{5}+\alpha_{6}y + \alpha_{7}+\alpha_{8}z \Bigr]+ G\eta,
\label{(4.47)}
\end{equation}
we arrive at the fundamental formula
\begin{equation}
f=f_{0}+Kf - GNf .
\label{(4.48)}
\end{equation}
At this stage, if the function $f$ has a 
Poincar\'e asymptotic expansion \cite{P1886}, 
which can be convergent or divergent and is written in the form 
\begin{equation}
f \sim \sum_{n=0}^{\infty}f_{n}, \; {f_{p}\over f_{0}} << 1 \; 
\forall p=1,2,...,\infty,
\label{(4.49)}
\end{equation}
we point out that (4.49) leads in turn to a Poincar\'e 
asymptotic expansion of the nonlinear term $Nf$ defined in Eq. (4.34) in the form
\begin{equation}
Nf \sim \sum_{n=0}^{\infty}A_{n}=A_{0}[f_{0}]+A_{1}[f_{0},f_{1}]+...
+A_{k}[f_{0},...,f_{k}]+...,
\label{(4.50)}
\end{equation}
where, by virtue of the formula
\begin{equation}
\log (1+\omega) \sim \omega-{\omega^{2}\over 2}+{\omega^{3}\over 3}+...
= \lim_{N \to + \infty} \sum_{k=0}^{N}(-1)^{k}{\omega^{k+1}\over k} \;
{\rm as} \; \omega \rightarrow 0,
\label{(4.51)}
\end{equation}
we can evaluate the Poincar\'e  
asymptotic expansion of squared logarithmic derivatives according to
\begin{eqnarray}
\left({1 \over f}{\partial f \over \partial x^{l}}\right)^{2}&=& \left({f_{,l}\over f}\right)^{2}
= \left \{ \left[\log f_{0} + \log \left(1+{f_{1}\over f_{0}}+{f_{2}\over f_{0}}+... \right)
\right]_{,l} \right \}^{2} 
\nonumber \\
& \sim & \left[{(f_{{0},l}+f_{{1},l})\over f_{0}}
-{f_{1}f_{{0},l}\over (f_{0})^{2}}
-{f_{1}f_{{1},l}\over (f_{0})^{2}}
+{(f_{1})^{2}f_{{0},l}\over (f_{0})^{3}}+... \right]^{2}.
\label{(4.52)}
\end{eqnarray}
In light of (4.34) and (4.52) we find
\begin{equation}
A_{0}[f_{0}]={3 \over 4f_{0}}\left[\sum_{l=1}^{3}t^{-2p_{l}}(f_{{0},l})^{2}
-(f_{{0},t})^{2}\right],
\label{(4.53)}
\end{equation}
\begin{equation}
A_{1}[f_{0},f_{1}]={3 \over 4f_{0}}\left(f_{1}\left(1-{f_{1}\over f_{0}}\right)\right)^{2}
\left \{ \sum_{l=1}^{3}t^{-2p_{l}}
\left[{f_{{1},l}\over f_{1}}
-{f_{{0},l}\over f_{0}}\right]^{2}
-\left[{f_{{1},t}\over f_{1}}
-{f_{{0},t}\over f_{0}}\right]^{2}\right \},
\label{(4.54)}
\end{equation}
plus a countable infinity of other formulas for $A_{2}[f_{0},f_{1},f_{2}]$,...
$A_{k}[f_{0},f_{1},...,f_{k}]$ ... . Note that, unlike the case of simpler nonlinearities
\cite{Adomian}, the functionals $A_{n}$ involve division by $f_{0},f_{1},...$.
The solution algorithm is now completely specified,
because Eq. (4.48) yields the recursive formulas \cite{Adomian}
\begin{equation}
f_{n+1}=Kf_{n}-GA_{n}[f_{0},...,f_{n}] \;
\forall n=0,1,...,\infty,
\label{(4.55)}
\end{equation}
and hence
\begin{equation}
\sum_{p=0}^{n}f_{p}=(I+K+...+K^{n})f_{0}-(I+K+...+K^{n-1})GA_{0}-...
-(I+...+K^{n-2})GA_{n-2}-GA_{n-1},
\label{(4.56)}
\end{equation}
where, by exploiting the partial sum of the geometric series, we find
\begin{equation}
f \sim \lim_{n \to \infty} \sum_{p=0}^{n}f_{p}
=\lim_{n \to \infty} \left[{(I-K^{n+1})\over (I-K)}f_{0}
-{(I-K^{n})\over (I-K)}GA_{0}-...-GA_{n-1}\right].
\label{(4.57)}
\end{equation}
Since the operator $K$ is built from the inverses of differential operators, it is a
pseudo-differential operator, and it remains to be seen whether, for sufficiently large values
of $n$, it only contributes to the terms $R_{j}(t)$ in the parametrix (4.4), so that we
only need the limit
\begin{equation}
\lim_{n \to \infty} \Bigr[(I-K)^{-1}(f_{0}-GA_{0})-...-GA_{n-1}\Bigr].
\label{(4.58)}
\end{equation}
The Adomian method we have used is well suited to go beyond weak nonlinearity and small perturbations,
but of course the nontrivial technical problem is whether the series for the unknown function $f$
is convergent, and also how fast. If it were necessary to consider hundreds of terms, the algorithm
would be of little practical utility. 

An interesting alternative, which cannot be ruled out at present, is instead the existence of
an asymptotic expansion of $f$ involving only finitely many terms, whose rigorous theory is described
in a monograph by Dieudonn\'e \cite{Dieu}. In such a case we might write 
\begin{equation}
f \sim f_{0}+f_{1}+f_{2}=(I+K+K^{2})f_{0}-(I+K)GA_{0}[f_{0}]
-GA_{1}[f_{0},f_{1}=Kf_{0}-GA_{0}[f_{0}]],
\label{(4.59)}
\end{equation}
which is fully computable by virtue of Eqs. (4.45)-(4.47) and (4.53)-(4.55).
We find it therefore encouraging that an exact solution
algorithm has been obtained for the scalar parametrix in Kasner spacetime.

Last, but not least, Eqs. (4.19) for the gradient of phase functions can be integrated
to find
\begin{eqnarray}
\varphi_{k}&=& \int (\alpha_{jk})^{-2}\psi_{0}{\rm d}t+\Phi_{0,k}(x,y,z,\xi_{1},\xi_{2},\xi_{3})
\nonumber \\
&=& \int (\alpha_{jk})^{-2}\psi_{1}{\rm d}x + \Phi_{1,k}(t,y,z,\xi_{1},\xi_{2},\xi_{3}) 
\nonumber \\
&=& \int (\alpha_{jk})^{-2}\psi_{2}{\rm d}y + \Phi_{2,k}(t,x,z,\xi_{1},\xi_{2},\xi_{3})
\nonumber \\
&=& \int (\alpha_{jk})^{-2}\psi_{3}{\rm d}z +\Phi_{3,k}(t,x,y,\xi_{1},\xi_{2},\xi_{3}),
\label{(4.60)}
\end{eqnarray}
bearing in mind that $\alpha_{jk}={1 \over \sqrt{t}}(f_{jk})^{1 \over 4}$ and Eq. (4.23), while 
the $\Phi$ functions may be fixed by demanding consistency with Eq. (4.5). This method leads to
the following formulas for the complete evaluation of phase functions:
\begin{equation}
\Phi_{0,k}(x,y,z,\xi_{1},\xi_{2},\xi_{3})=x \xi_{1}+y \xi_{2}+z \xi_{3}
- \lim_{t \to 0} \int (f_{jk})^{-{1 \over 2}}{\tilde \psi}_{0}{\rm d}t,
\label{(4.61)}
\end{equation}
\begin{equation}
\lim_{t \to 0} \Phi_{l,k}(t,X^{l},\xi_{1},\xi_{2},\xi_{3})=x \xi_{1}+y \xi_{2}+z \xi_{3}
- \lim_{t \to 0} \int (f_{jk})^{-{1 \over 2}}t^{2p_{l}}{\tilde \psi}_{l}{\rm d}x^{l},
\; \forall l=1,2,3,
\label{(4.62)}
\end{equation}
where $X^{l}$ denotes the triplet $x^{1}=x,x^{2}=y,x^{3}=z$ deprived of the $l$-th coordinate,
and no summation over $l$ is performed on the right-hand side.

\section{Concluding remarks}
 
The work in Ref. \cite{DEV14} succeeded in the difficult task of setting up a solution algorithm for
defining and solving self-dual gravity field equations to first order in the noncommutativity matrix.
However, precisely the first building block, i.e. the task of solving the scalar field equation in a
classical self-dual background was only briefly described. 

This incompleteness has been taken care of in the present paper for the case of
Kasner spacetime, first with a particular choice of Kasner parameters: $p_{1}=1,p_{2}=p_{3}=0$.    
The physics-oriented literature had devoted efforts to evaluating quantum propagators for a massive scalar 
field in the Kasner universe \cite{Nariai}, but the relevance for the classical wave equation of the
mathematical work in Refs. \cite{WW27,Garabedian,Oleinik,Menikoff,Tahara,Gar60} had not been appreciated,
to the best of our knowledge. As far as we know, our original results in Secs. III and IV are 
substantially new. We have indeed evaluated the bicharacteristics of Kasner spacetime in terms of
elliptic integrals of first, second and third kind, while the nonlinear system for obtaining 
amplitude and phase functions in the scalar parametrix\footnote{We note,
incidentally, that rediscovering the versatility of parametrices might lead to important progress
in canonical quantum gravity, since the work in Ref. \cite{DW1960} obtained 
diffeomorphism-invariant Poisson brackets on the space of observables, i.e. diff-invariant functionals 
of the metric, but this relied upon exact Green functions obeying advanced and retarded boundary
conditions, whereas the parametrix is what is strictly needed in the applications, and it might
prove more useful in defining and evaluating quantum commutators.} has been first mapped into Eqs.
(4.18)-(4.20), a set of equations that holds in any curved spacetime. 
Furthermore, the nonlinear equation (4.20) has been mapped into Eq. (4.31), and the
latter has been solved with the help of the Adomian method, arriving at Eqs. (4.53)-(4.59).

There is however still a lot of work to do, because the proof that the asymptotic expansion of
$f_{jk} \equiv ({\tilde \alpha}_{jk})^{4}$ is of the Poincar\'e   
type or, instead, only involves finitely many terms, might require new insight from asymptotic
and functional analysis. This adds evidence in favour of 
noncommutative gravity needing the whole apparatus of classical
mathematical physics for a proper solution of its field equations (see also the work in Ref.
\cite{Palia}, where Noether-symmetry methods have been used to evaluate the potential term
for a wave-type operator in Bianchi I spacetime).

\section*{Acknowledgments}
G. E. and E. D. G. are grateful to the Dipartimento di Fisica of Federico II University, Naples, for
hospitality and support.

\begin{appendix}

\section{Assessment of our wave equation and its solution}

As we know from Sec. V \cite{DEV14}, Eq. (1.3) is a particular case of the wave equation (3.1).
The operator $P$ in Eq. (3.1) is an example of what is called, in the mathematical literature,
a Fuchsian hyperbolic operator with weight $2$ with respect to $t$. In general, the weight is $m-k$,
and such Fuchsian hyperbolic operators read as (hereafter $(t,x)=(t,x^{1},...,x^{n}) \in
[0,T] \times {\bf R}^{n}$)
\begin{eqnarray}
P(t,x,\partial_{t},\partial_{x})&=& t^{k}\partial_{t}^{m}
+P_{1}(t,x,\partial_{x})t^{k-1}\partial_{t}^{m-1}+...+P_{k}(t,x,\partial_{x})
\partial_{t}^{m-k} \nonumber \\
&+& P_{k+1}(t,x,\partial_{x})\partial_{t}^{m-k-1}+...+P_{m}(t,x,\partial_{x}),
\label{(A1)}
\end{eqnarray}
subject to $10$ conditions stated in Ref. \cite{Tahara} which specify the relation between $k$ and
$m$, the form of the coefficients, hyperbolicity, quadratic form associated to the operator
(see below), estimates for principal part and lower order terms of the operator. When all these $10$
conditions hold, one can prove the following theorem \cite{Tahara}:
\vskip 0.3cm
\noindent
{\bf Theorem A1}. For any functions $u_{0}(x),...,u_{m-k-1}(x) \in C^{\infty}({\bf R}^{n})$
and $f(t,x) \in C^{\infty}([0,T] \times {\bf R}^{n})$, there exists a unique solution 
$u(t,x) \in C^{\infty}([0,T] \times {\bf R}^{n})$ such that
\begin{equation}
P(t,x,\partial_{t},\partial_{x}) u(t,x)=f(t,x) \; {\rm on} \; [0,T] \times {\bf R}^{n},
\label{(A2)}
\end{equation}
\begin{equation}
\left . \partial_{t}^{i}u(t,x) \right |_{t=0}=u_{i}(x) \; {\rm for} \;
0 \leq i \leq m-k-1,
\label{(A3)}
\end{equation}
and the solution has a finite propagation speed.

For the operator in Eq. (3.1), the quadratic form of the general theory, obtained by replacing
$$
{\partial \over \partial x_{j}} \rightarrow {\rm i} \xi_{j}
$$
in all spatial derivatives of second order, reads as
\begin{equation}
S(t,\xi)=\sum_{j=1}^{3}t^{-2p_{j}}(\xi_{j})^{2}.
\label{(A4)}
\end{equation}
According to Tahara, for Fuchsian hyperbolic operators, the quadratic form $S(t,\xi)$ leading to
Theorem A1 should be positive-definite as a function of $\xi$ for any $t>0$, with symmetric
coefficients of class $C^{1}$ on $[0,T]$, and such that
\begin{equation}
{\rm max}_{|\xi|=1} \left | {\partial \over \partial t} \log S(t,\xi) \right | 
={\rm O} \left({1 \over t}\right) \; {\rm as} \; t \rightarrow 0^{+}.
\label{(A5)}
\end{equation}
For the operator in Eq. (3.1) one finds indeed
\begin{equation}
{\partial \over \partial t} \log S(t,\xi)= -{2 \over t}
{\sum_{j=1}^{3}p_{j}t^{-2p_{j}}(\xi_{j})^{2} \over
\sum_{j=1}^{3}t^{-2p_{j}}(\xi_{j})^{2}} .
\label{(A6)}
\end{equation}
Thus, bearing in mind that, when the $p_{j}$ Kasner parameters are all nonvanishing, one of them is
negative and the other two are positive, one obtains (on defining $p \equiv {\rm max} 
\left \{ p_{j} \right \}$ for all $p_{j}>0$)
\begin{equation}
\left | {\partial \over \partial t} \log S(t,\xi) \right | \sim 
\left | {2 p \over t} \right | \; {\rm as} \; t \rightarrow 0^{+},
\label{(A7)}
\end{equation}
and hence condition (A5) of the general theory is fulfilled. This is also the case of the
operator in Eq. (1.3), for which
\begin{equation}
S(t,\xi)=t^{-2}(\xi_{1})^{2}+(\xi_{2})^{2}+(\xi_{3})^{2},
\label{(A8)}
\end{equation}
which implies that
\begin{equation}
\left | {\partial \over \partial t} \log S(t,\xi) \right |
= \left | {2 \over t} {t^{-2}(\xi_{1})^{2} \over t^{-2}(\xi_{1})^{2}+(\xi_{2})^{2}+(\xi_{3})^{2}}
\right |= {\rm O}\left({1 \over t}\right) \; {\rm as} \; t \rightarrow 0^{+}.
\label{(A9)}
\end{equation}
In other words, the hyperbolic equation studied in our paper can always rely upon the
Tahara theorem on the Cauchy problem. 

If instead we resort to the Garabedian technique of 
integration in the complex domain, strictly speaking,
we need to assume analytic coefficients \cite{Garabedian}, which is not fulfilled, for example, by
$b={1 \over t}$ in (2.19) if we replace $t$ by a complex $\tau=\tau_{1}+{\rm i} \tau_{2}$ and
want to consider also the value $\tau_{1}=\tau_{2}=0$. However, Ref. \cite{Garabedian} describes
the way out of this nontrivial technical difficulty. For this purpose, one considers first a more
complicated, inhomogeneous equation
\begin{equation}
L[u]=f
\label{(A10)}
\end{equation}
with analytic coefficients and analytic right-hand side, from which one can write down a direct analogue 
of the solution (2.20) in the form
\begin{equation}
u(t,x)=\lim_{\partial D \rightarrow T}
\left[ \int_{\partial D}B[u,{\cal P}]
+ \int_{D}(Pf -u M[{\cal P}]){\rm d}\tau \wedge {\rm d}y^{1}
\wedge {\rm d} y^{2} \wedge {\rm d} y^{3} \right],
\label{(A11)}
\end{equation}
where ${\cal P}(x,y)$ is called a {\it parametrix} (i.e. a distribution \cite{Friedlander}
that provides an approximate inverse) and is given by
\begin{equation}
{\cal P}(x,y)=\sum_{l=0}^{\nu}U_{l}(x,y)\sigma^{l-m}(x,y)
+\sum_{l=0}^{\mu}V_{l}(x,y)\sigma^{l}(x,y)\log \sigma(x,y),
\label{(A12)}
\end{equation}
in terms of the world function $\sigma(x,y)$ of Appendix B. The notation $\partial D \rightarrow T$ 
means that the manifold of integration $D$ is supposed to approach the real domain in such a way that it
folds around the characteristic conoid $\sigma=0$ without intersecting it. Equation (A11) defines a
Volterra integral equation for the solution of the Cauchy problem. It follows that $u$ varies
continuously with the derivatives of the coefficients of Eq. (A10). Similarly, the second partial
derivatives of $u$ depend continuously on the derivatives of the coefficients of a high enough order.
Thus, when they are {\it no longer analytic, we may replace these coefficients by polynomials
approximating an appropriate set of their derivatives} in order to establish the validity of (A11) in the
general case by passage to the limit. Note also that the integral equation (A11) has a meaning in the real
domain even where the partial differential equation (A10) is not analytic, since the construction
of the parametrix ${\cal P}(x,y)$ and of the world function $\sigma(x,y)$ only requires differentiability
of the coefficients of a sufficient order \cite{Garabedian}.

More precisely, for coefficients possessing partial derivatives of all orders, we introduce a polynomial
approximation that includes enough of these derivatives to ensure that the solution of the corresponding
approximate equation (A11) converges together with its second derivatives. The limit has therefore to be
a solution of the Cauchy problem associated with the more general coefficients, and must itself satisfy
the Volterra integral equation (A11).

\section{World function and fundamental solutions}

In his analysis of partial differential equations, Hadamard discovered the importance of the 
{\it world function} \cite{Hadamard, Ruse, Synge}, which can be defined as the square of the geodesic distance 
between two points with respect to the metric
\begin{equation}
g=\sum_{i,j=1}^{n} g_{ij}{\rm d}x^{i} \otimes {\rm d}x^{j}.
\label{(B1)}
\end{equation}
In the analysis of second-order linear partial differential equations
\begin{equation}
N[u]=\left[\sum_{i,j=1}^{n}a^{ij}{\partial^{2}\over \partial x^{i} \partial x^{j}}
+\sum_{i=1}^{n}b^{i}{\partial \over \partial x^{i}}+c \right]u=0,
\label{(B2)}
\end{equation}
the first-order nonlinear partial differential equation (cf. Eq. (3.5)) 
for the world function $\sigma(x,y)$ reads as \cite{Garabedian}
\begin{equation}
\sum_{i,j=1}^{n}a^{ij}{\partial \sigma(x,y) \over \partial x^{i}}
{\partial \sigma(x,y) \over \partial x^{j}}
=\sum_{i,j=1}^{n}a^{ij}{\partial \sigma(x,y) \over \partial y^{i}}
{\partial \sigma(x,y) \over \partial y^{j}}
=4 \sigma(x,y),
\label{(B3)}
\end{equation}
where the coefficients $a^{ij}$ are the same as those occurring in the definition of the operator
$N$ (this is naturally the case because the wave or Laplace operator can be always defined through
the metric, whose signature determines the hyperbolic or elliptic nature of the operator, as
we stressed in Sec. I). The world function can be used provided that 
the points $x$ and $y$ are so close to each other that no caustics occur.

A {\it fundamental solution} $S=S(x,y)$ of Eq. (B2) is a distribution \cite{Friedlander}, and  
can be defined to be \cite{Garabedian} a solution
of that equation in its dependence on $x=(x^{1},...,x^{n})$ possessing, at the parameter point 
$y=(y^{1},...,y^{n})$, a singularity characterized by the representation 
\begin{equation}
S(x,y)={U(x,y) \over (\sigma(x,y))^{m}}+V(x,y) \log(\sigma(x,y))+W(x,y),
\label{(B4)}
\end{equation}
where $U,V,W$ are supposed to be regular functions of $x$ in a neighbourhood of $y$, with
$U \not = 0$ at $x=y$, and where the exponent $m$ depends on the spacetime dimension $n$
according to $m={(n-2)\over 2}$. The sources of nonvanishing $V$ are either a mass term in
the operator $N$ \cite{DeWitt65} or a nonvanishing spacetime curvature \cite{Bimonte}. The
term $V \log(\sigma)$ plays an important role in the evaluation of the integral (2.20), as is
stressed in Sec. 6.4 of Ref. \cite{Garabedian}. 

In Kasner spacetime, the Hadamard Green function (B4) has been evaluated explicitly only
with the special choice of parameters $p_{1}=p_{2}=0,p_{3}=1$ in Ref. \cite{Nariai}.
In that case, direct integration of the geodesic equation (Appendix C)
yields eventually an exact formula for the Hadamard-Ruse-Synge 
world function in the form \cite{Nariai}
\begin{equation}
\sigma=t_{0}^{2}\Bigr(r_{\perp}^{2}-\tau^{2}-{\tau'}^{2}+2\tau \tau' \cosh(r_{3}) \Bigr),
\label{(B5)}
\end{equation}
having defined
\begin{equation}
\tau \equiv {t \over t_{0}}, \; \tau' \equiv {t' \over t_{0}},
\label{(B6)}
\end{equation}
\begin{equation}
r_{\perp} \equiv {\sqrt{(x_{1}-y_{1})^{2}+(x_{2}-y_{2})^{2}}\over t_{0}}, \;
r_{3} \equiv {(x_{3}-y_{3})\over t_{0}}.
\label{(B7)}
\end{equation} 
Following our remarks at the end of Sec. II, we expect that the choice of Kasner parameters
made in Sec. II would still lead to a formula like (B5) for the world function, but with
\begin{equation}
r_{\perp} \equiv {\sqrt{(x_{2}-y_{2})^{2}+(x_{3}-y_{3})^{2}}\over t_{0}}, \;
r_{1} \equiv {(x_{1}-y_{1})\over t_{0}}.
\label{(B8)}
\end{equation} 
However, as far as we know, the extension of these formulas to generic values of Kasner 
parameters is an open problem.

\section{World function of a Kasner spacetime}

The calculation in Ref. \cite{Nariai} is so enlightening and relevant for our purposes that it deserves
a brief summary. To begin, the geodesic equation in a Kasner spacetime with metric
\begin{equation}
g=-{\rm d}t \otimes {\rm d}t + \sum_{i=1}^{3}t^{2p_{i}}{\rm d}x^{i} \otimes {\rm d}x^{i}
\label{(C1)}
\end{equation}
is the following coupled system of nonlinear differential equations:
\begin{equation}
{{\rm d}^{2}t \over {\rm d}\lambda^{2}}+\sum_{i=1}^{3}p_{i}t^{2p_{i}-1}
\left({{\rm d}x^{i}\over {\rm d}\lambda}\right)^{2}=0,
\label{(C2)}
\end{equation}
\begin{equation}
{{\rm d}^{2}x^{i}\over {\rm d}\lambda^{2}}+{2 p_{i}\over t}{{\rm d}x^{i}\over {\rm d}\lambda}
{{\rm d}t \over {\rm d}\lambda}=0,
\label{(C3)}
\end{equation}
where $\lambda$ is the affine parameter of the geodesic. Equation (C3) can be solved for
$Y^{i} \equiv {{\rm d}x^{i}\over {\rm d}\lambda}$, because it yields
\begin{equation}
{{\rm d}\over {\rm d}\lambda} \log (Y^{i})={{\rm d}\over {\rm d}\lambda} 
\log (t^{-2p_{i}}),
\label{(C4)}
\end{equation}
which implies
\begin{equation}
{{\rm d}x^{i}\over {\rm d}\lambda}=n_{i}t^{-2p_{i}},
\label{(C5)}
\end{equation}
having denoted by $n_{1},n_{2},n_{3}$ three integration constants. The constancy along the geodesic of
the (pseudo-)norm squared $g_{\mu \nu}Y^{\mu}Y^{\nu}$, where 
$Y^{\mu} \equiv {{\rm d}x^{\mu}\over {\rm d}\lambda} (\mu=0,1,2,3)$, yields
($\varepsilon$ being negative (resp. positive) for timelike (resp. spacelike) geodesics)
\begin{equation}
\varepsilon=-\left({{\rm d}t \over {\rm d}\lambda}\right)^{2}
+g_{ij}{{\rm d}x^{i}\over {\rm d}\lambda} {{\rm d}x^{j}\over {\rm d}\lambda}
=-\left({{\rm d}t \over {\rm d}\lambda}\right)^{2}+\sum_{i=1}^{3}(n_{i})^{2}t^{-2p_{i}},
\label{(C6)}
\end{equation}
from which we obtain
\begin{equation}
{\rm d}\lambda={{\rm d}t \over \sqrt{\sum_{i=1}^{3}(n_{i})^{2}t^{-2p_{i}}-\varepsilon}}.
\label{(C7)}
\end{equation}
On the other hand, the world function is the square of the geodesic distance between the points 
$P'$ and $P$, say, i.e.
\begin{equation}
\sigma=\varepsilon L^{2}, \; L \equiv \int_{P'}^{P}{\rm d}\lambda
=\int_{\tau'}^{\tau}{{\rm d}t \over \sqrt{\sum_{i=1}^{3}(n_{i})^{2}t^{-2p_{i}}
-\varepsilon}}.
\label{(C8)}
\end{equation}
Moreover, following Ref. \cite{Nariai}, one defines
\begin{equation}
r_{i} \equiv (x^{i}-x'^{i})=n_{i}\int_{\tau'}^{\tau}
{t^{-2p_{i}}{\rm d}t \over \sqrt{\sum_{j=1}^{3}(n_{j})^{2}t^{-2p_{j}}
-\varepsilon}}.
\label{(C9)}
\end{equation}

Upon considering the particular choice $p_{1}=p_{2}=0,p_{3}=1$, and defining
\begin{equation}
N_{1,2}^{\varepsilon} \equiv (n_{1})^{2}+(n_{2})^{2}-\varepsilon ,
\label{(C10)}
\end{equation}
these formulae make it possible to re-express the integration constants in the form
\begin{equation}
n_{k} \equiv {r_{k} \over \sqrt{\tau^{2}N_{1,2}^{\varepsilon}+(n_{3})^{2}}
-\sqrt{\tau'^{2}N_{1,2}^{\varepsilon}+(n_{3})^{2}}} \; \forall k=1,2,
\label{(C11)}
\end{equation}
\begin{equation}
n_{3}={{\tau' {\rm e}^{-r_{3}} \sqrt{\tau^{2}N_{1,2}^{\varepsilon}+(n_{3})^{2}}
- \tau \sqrt{\tau'^{2}N_{1,2}^{\varepsilon}+(n_{3})^{2}}} \over
(\tau-\tau' {\rm e}^{-r_{3}})},
\label{(C12)}
\end{equation}
where $(r_{1})^{2}+(r_{2})^{2} \equiv r_{\perp}^{2}$. We can now square up the product
$n_{3}(\tau-\tau' {\rm e}^{-r_{3}})$ from (C12), finding eventually
\begin{equation}
(n_{3})^{2}=-\tau \tau' N_{1,2}^{\varepsilon}\cosh (r_{3})
+ \sqrt{\tau^{2} N_{1,2}^{\varepsilon}+(n_{3})^{2}}
\sqrt{\tau'^{2} N_{1,2}^{\varepsilon}+(n_{3})^{2}}.
\label{(C13)}
\end{equation}
On the other hand, the geodesic distance in (C8) becomes in our case 
\begin{equation}
L=\int_{\tau'}^{\tau}{{\rm d}t \over \sqrt{N_{1,2}^{\varepsilon}+{(n_{3})^{2}\over t^{2}}}}
={1 \over N_{1,2}^{\varepsilon}}\left[
\sqrt{\tau^{2}N_{1,2}^{\varepsilon}+(n_{3})^{2}}
-\sqrt{\tau'^{2}N_{1,2}^{\varepsilon}+(n_{3})^{2}} \right],
\label{(C14)}
\end{equation}
and if we square it up and then exploit (C13) we obtain
\begin{equation}
\sigma = \varepsilon {(\tau^{2}+\tau'^{2}-2 \tau \tau' \cosh(r_{3}))\over
N_{1,2}^{\varepsilon}},
\label{(C15)}
\end{equation}
because the terms involving products of square roots cancel each other. At this stage, we can
re-express the squares of $n_{1}$ and $n_{2}$ from (C11), i.e.
\begin{equation}
(n_{k})^{2}={(r_{k})^{2}N_{1,2}^{\varepsilon} \over 
(\tau^{2}+\tau'^{2}-2 \tau \tau' \cosh(r_{3}))} \; \forall k=1,2.
\label{(C16)}
\end{equation}
By virtue of (C15) and (C16), we find eventually the result (B5), where the role played by
$t_{0}$ in the formulas has been made explicit.

\end{appendix}

\end{document}